\documentclass[aps,twocolumn]{revtex4}
\usepackage{graphicx}
\usepackage[dvips]{epsfig}
\begin{document}
\title{Raman scattering on phonon--plasmon coupled modes in  magnetic fields}

\author{L. A. Falkovsky}
\affiliation{Landau Institute for Theoretical Physics, 119337 Moscow, Russia}
\bigskip
\bigskip

\begin{abstract}
Raman scattering  on  phonon--plasmon coupled modes
in high
magnetic fields is considered theoretically.
The calculations of the dielectric function
 were performed in the
long-wave approximation for the semiclassical and ultra-quantum
magnetic fields taking into account the electron damping and
intrinsic lifetime of optical phonons. The  Raman scattering has
resonances at the frequencies of coupled modes as well as at
multiples of the cyclotron frequency. The dependence of the Raman
cross section on the carrier concentration is analyzed.
\end{abstract}
\maketitle

\section{Introduction}

 Phonon-plasmon coupled modes arise as a result of the interaction
of optical phonons with charge carriers in semiconductors. These
modes  are usually observed in the absence of a magnetic field,
for instance, in  Raman scattering \cite{ACP}. Interest
in these modes is related not only to applications in optics
but also to
principal
questions of condensed-state physics, in particular, to the
possibility of determining the magnitude of  electron-phonon
interaction with their help. The first experimental
results for the Raman scattering on phonon-plasmon coupled modes
in magnetic fields were recently obtained
\cite{WPS}, but only the  preliminary theoretical paper \cite{Fph}
has been published  up to now.

The longitudinal optical phonons interact strongly with free
carriers because of an electric field accompanied in the lattice
vibrations.
Let us estimate  the  parameters of the corresponding
electron-phonon system. The frequency of an emitted or absorbed
optical phonon $\omega$ is usually equal to several hundreds of
degrees, for example, this is 35 mV in GaAs.  If the light scattering
is excited by the
laser frequency $\omega_i=1.5$ eV, the momentum transfer  $k$
 from the incident
radiation has the order of $\omega_i/c\sim 10^5$ cm$^{-1}$. Then, the
conditions $\omega/c\ll k\ll \omega/v_F$ are fulfilled, where the
Fermi velocity is $v_F\sim 0.5\times 10^8$ cm/c in the case when
the statistics of carriers is degenerate. The left-hand side of
the above inequality means that  the electric field associated
with the phonon vibration of  frequency $\omega$ is really
static and therefore the electric field is longitudinal
with respect to the wavevector ${\bf k}$. The right-hand side of the
inequality allows the electron contribution to the dielectric function
$\varepsilon({\bf k}, \omega)$ to be calculated using a series
expansion in terms of the dispersion parameter  $kv/\omega$.
We shall see that  at
least the lowest-order correction should be held,
because it has a resonant character. As for the lattice
contribution to $\varepsilon({\bf k}, \omega)$, it can be taken at
$k=0$, since the phonon dispersion is negligible  for such
the momentum.

The effect of magnetic field on optical phonons is most essential
if  the cyclotron frequency of carriers $\omega_c$
has the order of the optical phonon frequency. This means that
the magnetic field must reach 20 T for effective carrier mass of
0.063 $m_0$ in the same GaAs. Changing the carrier
concentration, we can observe both the semiclassical regime, when
the cyclotron frequency is much less than the Fermi energy
($\omega_c\ll\varepsilon_F$), and the ultra-quantum regime, when
$\omega_c > \varepsilon_F$.

Two circumstances should be noted. Firstly, both the frequency of
the coupled modes and the width of the corresponding resonance are
of interest in the Raman light scattering.
Therefore, the contribution of carriers to the dielectric
susceptibility must be calculated with regard to their damping.
The spatial dispersion of  susceptibility in a magnetic field
under these conditions has not been calculated so far. Secondly,
in addition to the Coulomb interaction of carriers with
the longitudinal phonon vibrations, which is taken into account by the
dielectric function, there exists a deformation interaction
with both the LO and TO modes, which bears
the  Fr\"ohlich name in theoretical works. It arises because of the
nonadiabaticity  of  electron-phonon systems and leads to a
certain (as small as the nonadiabaticity  parameter)
renormalization of phonon frequencies. In the absence of a
magnetic field, this renormalization has been recently considered
in \cite{AS} - \cite {Fal}, and it will not be taken into account
here.

The structure of the paper is the following. A short outline of
the  theory of  Raman scattering on the phonon-plasmon coupled mode
  is give in Sec. II. Then the dielectric function of
the system in
 semiclassical and ultra-quantum
regimes is evaluated in Secs. III and IV. The theoretical Raman
spectra are presented for the various magnetic fields and
carrier concentrations in Sec. V. Finally, the conclusions are
summarized in Sec. VI.

\section{Inelastic light scattering on phonon--plasmon coupled modes}

Let us consider the Raman scattering on optical vibrations
in a polar lattice
with free carriers. We  use the notations
${ b}_j$
for the  phonon displacements of the branch $j$.
The subscript $j$ denotes the various phonon modes: longitudinal
or transverse ones. More precisely, the subscript $j$ indicates
the different phonon representations which can be degenerate. The
 transformation properties of the coupling constants $g_{j}$
 are determined by this representation.
The LO vibrations are accompanied by the
internal electric field ${\bf E}( {\bf r}, t)$. We consider the phonon
displacements and  electric field as classical variables
because the magnetic field affects quantum-mechanically
only  the free carriers.

The effective Hamiltonian describing the inelastic light scattering
on phonon--plasmon modes can be written as
 \begin{equation} \label{hami}
{\cal H} =
 {e^{2} \over m c^{2}}
 \int d^{3}r\,
 {\cal N}({\bf r},t)U({\bf r},t),
 \end{equation}
where
\begin{equation} \label{1a}
{\cal N}({\bf r},t) =
g_{j}  b_j({\bf r},t)+
g_E E({\bf r},t)
\end{equation}
is a linear form of variables
$b_j$ and $E$. We do not include the pure electronic Raman scattering
by free carriers, which has been
considered in many papers (see, e.g. \cite {N}, \cite{Mi}).

The notation $U({\bf r}, t)$ is introduced for a product of the
vector-potentials of incident  and scattered  photons:
$$ A^{(i)}({\bf r},t) A^{(s)}({\bf r},t)=
U({\bf r}, t)=
\exp[i({\bf k}\cdot{\bf r}- \omega t)]
U({\bf k}, \omega),$$
where the momentum and frequency transfers
${\bf k} = {\bf k}^{(i)} - {\bf k}^{(s)}$,
$\omega = \omega^{(i)} - \omega^{(s)}$. The polarization vectors
$e^{(i)}_{\alpha }$ and $e^{(s)}_{\beta}$
of the incident  and scattered  photons
   are included in the coupling constants
   as well as the polarization vector of
   ${\bf E}({\bf r}, t)$ is included in the coupling constant $g_E$.
The estimation of the coupling constants gives $g_j\sim /a^4$,
$g_{E}\sim 1/ea$
where $a$ is the lattice parameter.

If we evaluate
the generalized  susceptibility $\chi ({\bf k}, \omega )$
 in the linear response to the
 force $U({\bf k} , \omega)$
 \begin{equation} \label{11}
{\cal N}({\bf k},\omega)
 \,= - \chi ({\bf k}, \omega )
U({\bf k} , \omega),
\end{equation}
we  obtain the  Raman cross-section
\begin{eqnarray} \nonumber
\frac{d\sigma}{d\omega^{(s)} d\Omega^{(s)}}=
\frac{ 2k_{z}^{(s)} \omega^{(s)}}{\pi c(1 - \exp(-\omega /T)}
\left( \frac{2e^{2}}
{ c \hbar m \omega^{(i)}}\right)^{2}\\
\times|U({\bf k},\omega)|^2 {\rm Im}\,
\chi ({\bf k}, \omega ),
\label{8}
\end{eqnarray}
where $k_{z}^{(s)}$ is the normal to the sample surface component
of the scattered wave vector in vacuum.

One note should be made here. Of course, any sample has  the surface.
The surface effects in the Raman
scattering was considered in the work \cite{FM} and they are  omitted in the
derivation of Eq. (\ref{8}). Furthermore, the incident and scattered fields
do not penetrate into the bulk due to the skin effect. For the optical range
of the incident light, we have the normal skin-effect conditions. Then we
 integrate in Eq. (\ref{8}) the distribution $|U({\bf k},\omega)|^2$
over the normal component $k_z$. As shown in the paper \cite{FM},
 the integration of  $|U({\bf k},\omega)|^2$  gives a factor $1/\zeta_2$,
 where $\zeta_2$ is expressed in terms of
the wave-vector components  inside semiconductor:
$\zeta_2= {\rm Im} (k_z^{(i)}+
k_z^{(s)})$. The Raman cross section (\ref{8}) obtained is dimensionless.
It represents a ratio
of the inelastic scattered light energy  to the incident energy.

The  equation of motion for the LO mode $b_{\rm LO}$ has the
form
\begin{equation}
\label{oeq}
(\omega_{\rm TO}^2-\omega ^{2})b_{\rm LO}({\bf k},\omega)
=\frac{Z}{M'}E({\bf k},\omega)-\frac{g_{\rm TO}U({\bf k},
\omega)}{M'N},
\end{equation}
where $N$ is the number of unit cells in 1 cm$^3$,
 $M'$ is the reduced mass of the unit cell, and $Z$ is the effective
 ionic charge.
 Notice that the optical phonons always have the so-called natural
width $\Gamma\sim \omega_0\sqrt{m/M}$. The natural width results from
decay processes into two or more acoustic and optical phonons. In the final
expressions, we will substitute
$\omega_{\rm TO}^2-\omega^2\rightarrow\omega_{\rm TO}^2-i\omega\Gamma-\omega^2$.
The equation (\ref{oeq}) is applied as well to  the transverse
 phonons, but the electric field has to be neglected
 in the case of the TO phonons.

 The electric field ${\bf E}({\bf r}, t)$ can be obtained from
  the Poisson equation ${\rm div}{\bf D}=0$.
There are several contributions in the  induction $ D$:
(1) the polarization $\alpha E({\bf r},t)$ of the filled
electron bands, (2) the lattice polarization
$NZb_{\rm LO}({\bf r},t)$,
(3) the contribution of free carrier density
$\rho=-{\rm div}{\bf P}_e$, and (4) the term
$P= -\partial{\cal H}/\partial E= - g_EU$  explicitly results
from the Hamiltonian, Eqs. (\ref{hami}), (\ref{1a}).
Collecting all these terms into the Poisson
equation, we find
\begin{eqnarray}\label{pe}\nonumber
\varepsilon_{\infty}E({\bf k},\omega)
+4\pi NZb_{\rm LO}({\bf k},\omega)\\
+\frac{4\pi ie}{k}
\rho({\bf k},\omega) -4\pi g_EU({\bf k},\omega)=0,
\end{eqnarray}
where the high-frequency dielectric constant
 $\varepsilon_{\infty}=1+4\pi\alpha$.

We shall calculate the carrier density $\rho({\bf k},\omega)$ in
the following sections and find the electron contribution in
the dielectric function
$\varepsilon_e({\bf k},\omega)=
\varepsilon_{\infty}+4\pi ie\rho({\bf k},\omega)/kE$.
Then Eq. (\ref{pe}) takes the form
\begin{equation}\label{pe1}
\varepsilon_{e}E({\bf k},\omega)
+4\pi NZb_{\rm LO}({\bf k},\omega)
 -4\pi g_EU({\bf k},\omega)=0.
\end{equation}
Solving Eqs. (\ref{oeq}), (\ref{pe1}) and using Eq. (\ref{1a}),
we find ${\cal N}({\bf k,}\omega)$
 and obtain the susceptibility defined by Eq. (\ref{11}):
\begin{equation}\label{suc}
\chi({\bf k}, \omega)=
4\pi g_{E}^2\frac{\varepsilon_e({\bf k}, \omega)
C^2\omega^4_{\rm TO}/\varepsilon_{\infty} \omega^2_{pi}
-\Delta
-2C \omega_{\rm TO}}
{\varepsilon_e({\bf k}, \omega)\Delta+\varepsilon_{\infty}\omega^2_{pi}},
\end{equation}
where we set
$\Delta=\omega_{\rm TO}^2-\omega^2-i\omega\Gamma$,
the  plasma frequency  of ions
$\omega_{pi}^2=4\pi NZ^2/\varepsilon_{\infty}M'$,
and the Faust-Henry coefficient $C=g_{\rm TO}Z/g_EM'\omega_{\rm TO}^2$.

We see from Eq. (\ref{suc}), that the poles of the general
susceptibility coincide with the zeros of the total dielectric
function
\begin{equation} \label{dft}
 \varepsilon({\bf k}, \omega) = \varepsilon_e({\bf k}, \omega)+
 \frac{\varepsilon_{\infty}\omega^2_{pi}}
 {\omega_{\rm TO}^2-\omega^2-i\omega\Gamma}.
\end{equation}
Therefore, the peaks of the Raman cross section give the
frequencies of coupled modes determined by the condition
$\varepsilon({\bf k}, \omega)=0$.

In order to calculate the generalized susceptibility, Eq. (\ref{suc}),
we must find the dielectric function $\varepsilon_e({\bf k}, \omega)$,
 i.e. the free carrier density $\rho({\bf k},\omega)$.

\section{The dielectric function in semiclassical regime}

Consider the geometry when the magnetic-field effect is most
important: the magnetic field ${\bf H}$ is directed along the axis $z$,
whereas the wavevector ${\bf k}$ and the corresponding  electric field
${\bf E}$ are directed  in the perpendicular direction $x$.
In semiclassical conditions $\omega_c\ll\varepsilon_F$,
we can use the Boltzmann equation
\begin{equation} \label{be}
-i(\omega - k  v_x) f+\omega_c\frac{d f}{d\varphi}=
 v_x-( f-< f>)/\tau,
\end{equation}
written in the $\tau-$approximation within the  momentum variables:  the
 component $p_z$ along the magnetic field, the angle  $\varphi$
 of rotation around the magnetic field, and the energy
 $\varepsilon$. The element of phase volume in these variables
 is $md\varepsilon dp_zd\varphi$.   For a
quadratic electron spectrum, we can assume, for example,
that $v_x=v_{\bot}(p_z)\sin\varphi$.
The factor
 $e{\bf E}\delta (\varepsilon-\varepsilon_F)$,
containing the charge, the electric field, and the Dirac function,
is separated out of the distribution function $ f$.
Angle brackets designate
the average over the Fermi surface
$$ <...>=\int (...)m dp_z d\varphi /\int m dp_z d\varphi .$$
The term with averaging in the
Boltzmann equation is necessary for the fulfillment of the
conservation law of the electric charge.

We assume $kv/\omega\ll 1$ and expand
the solution to Eq. (\ref{be})
in powers of this parameter  to the second order
$f=f_0+f_1+f_2$. Because $\langle v_x \rangle$ vanishes in
integrating over $\varphi$, it is seen
from Eq. (\ref{be}) that $\langle f_0\rangle=0$. Therefore, the system of
equations for $f_0, f_1$, and $f_2$ takes the form
\begin{eqnarray}\label{sb}\nonumber
-i\omega^*f_0+\omega_c\frac{df_0}{d\varphi}=
v_x,\\
-i\omega^*f_j+\omega_c\frac{df_j}{d\varphi}=
\frac{<f_j>}{\tau}-i k v_x f_{j-1}, \nonumber
\end{eqnarray}
where $j=1,2$ and $\omega^{*}=\omega + i/\tau$.  The zeroth- and
first-order approximation, which have to be periodic
in the angle $\varphi$, are readily found
\begin{eqnarray}\label{sbe}\nonumber
f_0=\int^{\varphi}_{-\infty}v_x\exp{\left[-i\omega^*(\varphi-\varphi')/\omega_c\right]}=\\
\frac{-v_{\perp}}{2}\left[(\omega_c-\omega^*)^{-1}e^{i\varphi}+
(\omega_c+\omega^*)^{-1}e^{-i\varphi}\right]
\end{eqnarray}

\begin{equation} \label{feq1}
f_1=\int^{\varphi}_{-\infty}(<f_1>/\tau)-ikv_xf_0)
\exp{\left[-i\omega^*(\varphi-\varphi')/\omega_c\right]},
\end{equation}
First, $\langle
f_1\rangle$ and, then, $f_1$ can be found by averaging both the
sides of the last equation. We obtain
\begin{eqnarray} \label{fe1}\nonumber
f_1=\frac{-ikv_{\perp}^2}{4(\omega_c^2-\omega^{*2})}
\left[\frac{4i}{3\omega\tau}+2\frac{v_{\perp}^2}{v_0^2}\right.\\
\left.+\frac{(\omega_c+\omega^*)v_{\perp}^2}{(2\omega_c-\omega^*)v_F^2}
e^{2i\varphi}
+\frac{(\omega_c-\omega^*)v_{\perp}^2}{(2\omega_c+\omega^*)v_F^2}
e^{-2i\varphi}\right].
\end{eqnarray}

For the second-order approximation, an expression similar to
$f_1$, Eq. (\ref{feq1}) is obtained. The difference is connected with the fact
that $f_2$ is an odd function of velocity, and, therefore, the
mean value $\langle f_2\rangle$ vanishes. It can be found that
\begin{eqnarray}\label{f2}\nonumber
f_2=\frac{k^2v_{\perp}v_F^2}{8(\omega_c^2-\omega^{*2})}
\left[\left(\frac{4i}{3\omega\tau}+2\frac{v_{\perp}^2}{v_0^2}\right)
\left( \frac{e^{i\varphi}}{\omega_c-
\omega^*}+\frac{e^{-i\varphi}}{\omega_c+\omega^*}\right)\right.\\ 
+
\frac{(\omega_c+\omega^*)v_{\perp}^2}{(2\omega_c-\omega^*)v_F^2}
\left( \frac{e^{3i\varphi}}{3\omega_c-\omega^*}
-
\left.\frac{e^{i\varphi}}{\omega_c-\omega^*}\right)\right.\\ \nonumber
\left.-\frac{(\omega_c-\omega^*)v_{\perp}^2}{(2\omega_c+\omega^*)v_F^2}
\left( \frac{e^{-i\varphi}}{\omega_c+\omega^*}-
\frac{e^{-3i\varphi}}{3\omega_c+\omega^*}\right)\right].
\end{eqnarray}

It is simpler to calculate the current $j({\bf k}, \omega)$
instead of the
electric charge, using the conservation law
$\rho({\bf k}, \omega)=k j({\bf k}, \omega)/\omega$.
The contribution of carrier into the conductivity is
\begin{equation}\label{con}
\sigma=e^2 \int v_x f m\frac{2 d\varphi dp_z}{(2\pi \hbar)^3}.
\end{equation}
Because the first-order approximation $f_1$, Eq. (\ref{fe1}) is even with
respect to velocity, it makes no contribution to the conductivity.
Using the zeroth-order, Eq. (\ref{sbe}) and the second-order, Eq. (\ref{f2})
approximations, we find the carrier  conductivity, Eq.
(\ref{con}) and the susceptibility $4\pi i\sigma/\omega$.
 With regard to the  contribution of filled bands, the electron
susceptibility  can be written as
\begin{eqnarray} \label{dp}
\varepsilon_e(\omega,k)=\varepsilon_{\infty}\nonumber
\left\{1-
\frac{\omega_{pe}^2\omega^*}{\omega(\omega^{*2}-\omega_c^2)}\right.\\ -
\left.\frac{3\omega_{pe}^2\omega^*k^2v_F^2}{5\omega(\omega^{*2}-\omega_c^2)^2}
\left(1+\frac{5i}{9\omega\tau}+
\frac{3\omega_c^2}{\omega^{*2}-4\omega_c^2}\right)
\right\}.
\end{eqnarray}
The expression obtained remains valid both in the absence of a
magnetic field $(\omega_c=0)$ and in the collisionless limit
$(\tau = \infty)$. If both these conditions are fulfilled,
Eq. (\ref{dp})
converts to the known expression with the true coefficient 3/5.
This coefficient, as well as the others in Eq. (\ref{dp}),
was calculated
here for the quadratic electron spectrum; however, the dependence
itself on the frequency $\omega$, magnetic field, and damping is
 retained in the general case.
It is easy to see that the highest resonances $n\omega_c$ contribute
to the dielectric function the terms of the order $(kv_F/\omega_c)^n$
with the even $n$.

The dielectric function of a nondegenerate electron
plasma was obtained in Ref. \cite{PWT}
\begin{equation} \label{pwt}
 \varepsilon(\omega,k) =
 1-\frac{\omega_{pe}^2\omega^*}{\omega}
 \left[\frac{1-\lambda}{\omega^{*2}-\omega_c^2}+
 \frac{\lambda}{\omega^{*2}-4\omega_c^2}\right],
 \end{equation}
where $\lambda = (kv_{th}/\omega_c)^2$, and $v_{th}$ is the
thermal velocity of electrons. At $\tau = \infty$ and $\lambda =
(kv_F/\omega_c)^2/5$, it coincides with the electronic term in Eq.
(\ref{dp}).

\section{Ultra-quantum magnetic fields}

Now we consider the very high magnetic field when only one lowest
electron level is occupied. In order to calculate the dielectric
function, we use the expansion of the electron density
\begin{equation} \label{eld}
\sum_{s,s'}\psi^{\ast}_s({\bf r})\psi_{s'}({\bf r})a^{+}_sa_{s'}
\end{equation}
in terms of the eigenfunctions
$$\psi_{s'}({\bf r})=\chi_n
(x-cp_y/eH)e^{i(p_z z+p_y y)/\hbar},$$
where $s$ is the total set of the quantum numbers $s=\{n,p_y,p_z\}$
 and $\chi_n$ can be expressed in  terms of
the Hermite polynomials. The  magnetic field is chosen in the
$z-$direction.

Using the equation of motion for the operator $a^{+}_sa_{s'}$ in the
weak  field
$\phi({\bf r})$, we find the linear response
\begin{equation} \label{rp}
\langle 0|a^{+}_sa_{s'}|0\rangle=\frac{e\phi_{s',s}(f_s-f_{s'})}
{\hbar\omega-\varepsilon _{s'}+\varepsilon_{s}} ,
\end{equation}
where $\phi_{s',s}$ is the matrix element of the potential $\phi({\bf
r})$, $f_s$ is the Fermi distribution function depending on
$\varepsilon_{s}$ and the average is taken over the ground state.
Inserting Eq. (\ref{rp}) into Eq. (\ref{eld}), averaging and taking
the Fourier transform with respect ${\bf r}$, we obtain
the induced charge and then the electronic dielectric function
\begin{eqnarray} \label{qdf} \nonumber
\varepsilon_e({\bf k},\omega)=\varepsilon_{\infty}-\frac{4\pi e^2}{k^2}
\sum_{n,n'}\int\frac{dxdp_ydp_z}{(2\pi\hbar)^2}e^{ik_xx}\\ 
\times \chi_n(-cp_y/eH)\chi_n(x-cp_y/eH) \\ \nonumber
\times \chi_{n'}(-cp'_y/eH)\chi_{n'}(x-cp'_y/eH)
\frac{(f_s-f_{s'})}
{\hbar\omega-\varepsilon _{s'}+\varepsilon_{s}},
\end{eqnarray}
where $p'_{y}=p_{y}+k_{y}$,
$\varepsilon_{s}=(n+1/2\pm g/4)\hbar\omega_c+p_z^2/2m^*$,
$\varepsilon_{s'}=(n'+1/2\pm g/4)\hbar\omega_c+p_z^2/2m^*$.
  In the following,
 we assume that the g-factor  equals  2 as for the free
 electron.

Let us introduce the new variables: $x-cp_y/eH\rightarrow x,
-cp_y/eH\rightarrow x'$. Then the integral (\ref{qdf}) takes the
form:
\begin{eqnarray} \label{xdf}\nonumber
\varepsilon_e({\bf k},\omega)=\varepsilon_{\infty}-\frac{4\pi e^3H}{ck^2}
\sum_{n,n'}\int\frac{dxdx'dp_z}{(2\pi\hbar)^2}
e^{ik_x(x-x')}\\ \nonumber \times 
\chi_n(x)\chi_n(x') 
\chi_{n'}(x-ck_y/eH)\chi_{n'}(x'-ck_y/eH)\\  \times 
\frac{(f_s-f_{s'})}
{\hbar\omega-\varepsilon _{s'}+\varepsilon_{s}}.
\end{eqnarray}
Because ${\bf H}$ is
perpendicular  to ${\bf k}$, we can choose $k_y=k_z=0$.
Then we get under the integral
(\ref{xdf})  the  matrix element
$I_{n,n'}=(e^{ik_xx})_{n,n'}$ squared.

The wave function $\chi_n(x)$
changes over the magnetic length $a_{H}=\sqrt{\hbar/m^*\omega_c}$.
We have  $k\ll 1/a_H$ for the high magnetic fields. Therefore
we can evaluate the integral $I_{n,n'}$ expanding the exponent
 in  powers of $k_x$ and using the known matrix elements
 $x_{n,n-1}=x_{n-1,n}=\sqrt{n\hbar/2m^*\omega_c}$.
Due to the Fermi factors
$f_s, f_{s'}$, the integrand in Eq. (\ref{xdf})
does not vanish only if   $n\ne n'$
and one of the numbers $n, n'$ have to be zero in the ultra-quantum limit.
 Then we are interested in
the off-diagonal matrix elements $I_{n,n'}$.
Under these conditions, we obtain  to the forth order in $k$:
\begin{eqnarray}\label{me}\nonumber
|I_{n,n'}|^2=\frac{\hbar k^2}{2m^*\omega_c}(n+1)\delta_{n',
n+1}\\ \nonumber -
\left(\frac{\hbar k^2}{2m^*\omega_c}\right)^2
\left[\frac{1}{3} (n+1)(2n+3)\delta_{n',n+1}\right.\\  -
\left.\frac{1}{4} (n+1)(n+2)\delta_{n',n+2}\right]+(n\leftrightarrow n').
\end{eqnarray}

Substituting Eq. (\ref{me}) into Eq. (\ref{xdf}),
taking terms with $n=0$ or $n'=0$, and integrating with respect
$p_z$, we find the electron contribution into the dielectric
function in the ultra-quantum limit
\begin{eqnarray} \label{fdf} \nonumber
\varepsilon_e({\bf k},\omega)=\varepsilon_{\infty}-
\frac{ 2e^2p_F\omega_c}{\pi\hbar^2}\\ \times
\left(\frac{1}{\omega^2-\omega_c^2}
-\frac{(ka_H)^2/2}{\omega^2-\omega_c^2}+
\frac{(ka_H)^2/2}{\omega^2-4\omega_c^2}\right),
\end{eqnarray}
where $p_F$ is the Fermi momentum in the magnetic field. In order
to find this value, we assume that the electron concentration
$$ \sum_{s}f_s|\chi_{n}(x-cp_y/eH)|^2$$
 is fixed by the defect concentration $n_0$. In the ultra-quantum limit,
 all the electrons are on the lowest level with $n=0$. Integrating
 with respect $p_y$ and $p_z$, we get
\begin{equation}\label{en}
n_0=2m^*\omega_cp_F/(2\pi\hbar)^2
\end{equation}
and Eq. (\ref{fdf}) becomes
\begin{eqnarray} \label{fd}
\varepsilon_e({\bf k},\omega)=\varepsilon_{\infty}-
\varepsilon_{\infty}\omega_{pe}^2\\ \nonumber\times 
\left(\frac{1}{\omega^2-\omega_c^2}
-\frac{(ka_H)^2/2}{\omega^2-\omega_c^2}+
\frac{(ka_H)^2/2}{\omega^2-4\omega_c^2}\right),
\end{eqnarray}
where $\omega_{pe}^2=4\pi n_0e^2/m^*\varepsilon_{\infty}$.
Therefore Eq. (\ref{fd}) coincides  with Eq. (\ref{dp}),
if  $1/\tau=0$ and
$v_F^2=5\hbar\omega_c/2m^*$. We see, that
the parameter $\lambda=(kv_F/\omega_c)^2/5\sim (k/H)^2$ in
the semiclassical case, Eq. (\ref{pwt}),
whereas the corresponding parameter in the ultra-quantum  limit,
Eq. (\ref{fd}), is
$\lambda '=(ka_H)^2/2\sim k^2/H$.

The effect of collisions was not taken into account in Eq.
(\ref{fd}). It can be done with the help of the simple phenomenology
similar to the paper \cite{Me}. But for the case of our interest
$1/\tau\ll \omega$, when the coupled modes are observed, we can
merely replace $\omega$ by $\omega+1/\tau$ in Eq. (\ref{fd}).

\section{Theoretical Raman spectra}

The zeros of the dielectric function
$\varepsilon({\bf k},\omega)$, Eq. (\ref{dft}) give the frequencies of the
longitudinal modes.
With the neglect of the electron  $\tau^{-1}$ and
phonon $\Gamma$ damping, the dielectric function, Eqs.
(\ref{dp}) and (\ref{fd}),
  is real and the frequencies of the
corresponding vibrations are also real. In order to determine
these frequencies, one has to solve a cubic equation.

\begin{figure}
\begin{center}
\epsfxsize=80mm
\epsfysize=64mm
\epsfbox{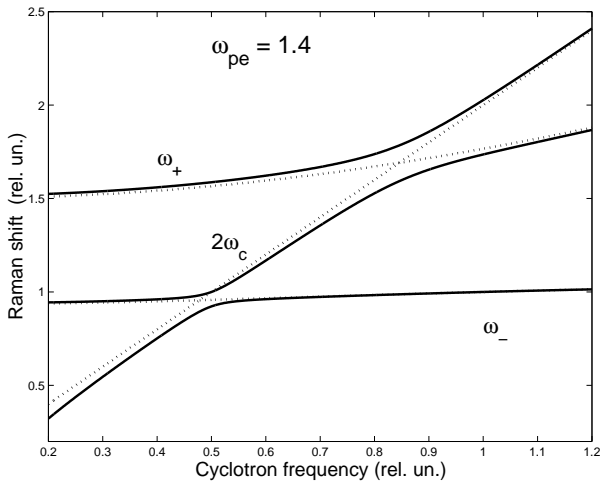}
\end{center}
\caption{
Frequencies  of the phonon-plasmon
coupled modes as functions of  the  carrier cyclotron
frequency  for the carrier
plasma frequency
$\omega_{pe}$ = 1.4 and $\omega_{\rm LO}=1.1$; all frequencies are
in units $\omega_{\rm TO}$. The solid lines are obtain in
solution of the cubic equation (see text),
the dotted lines show the results ignoring the $k-$dispersion.}
\end{figure}

To a good approximation
 one  can first omit the small term with $k^{2}$ and
obtain two eigenfrequencies $\omega_{\pm}$
\begin{eqnarray} \nonumber
\omega_{\pm}^2=\frac{1}{2}(\omega_c^2+\omega_{\rm TO}^2+
\omega_{pi}^2+\omega_{pe}^2)\\
\pm\frac{1}{2}[(\omega_c^2-\omega_{\rm TO}^2-
\omega_{pi}^2+\omega_{pe}^2)^2+
4\omega_{pi}^2\omega_{pe}^2]^{1/2}
\end{eqnarray}
by solving the corresponding quadratic
equation.
The third frequency is found in the vicinity of the pole
$\omega = 2\omega_c$ of the omitted term. This approximation is shown
in dotted  lines in Figs. 1 and 2.

\begin{figure}
\begin{center}
\epsfxsize=80mm
\epsfysize=64mm
\epsfbox{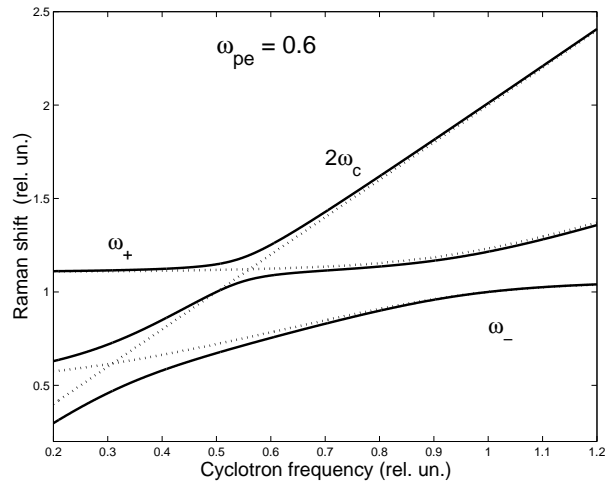}
\end{center}
\caption{The same as in Fig. 1 but for
$\omega_{pe}=0.6.$ }
\end{figure}
The numerical results including the $k-$dispersion  are shown in solid lines.
Now we must distinguish  the semiclassical   and
ultra-quantum regimes. In the last case, the condition
$\hbar\omega_c>\varepsilon_F$ is fulfilled and the Fermi energy
$\varepsilon_F$ depends on the magnetic field according to Eq. (\ref{en}).
Then the condition of ultra-quantum regime is
rewritten as $(2eH/\hbar c)^{3/2}>(2\pi)^2n_0$ for the fixed
carrier concentration $n_0$. For instance, the ultra-quantum regime is
realized for the concentration $n_0=10^{17}$ cm$^{-3}$  at $H>7.8$ T and
for $n_0=10^{18}$ cm$^{-3}$   at $H>36 $ T.
We suppose  that  the semiclassical condition, Eq. (\ref{dp})
 is obeyed in the  interval of magnetic fields displaced in Fig.
 1, whereas the ultra-quantum regime, Eq. (\ref{fd}) is assumed in Fig. 2.

\begin{figure}
\begin{center}
\epsfxsize=80mm
\epsfysize=64mm
\epsfbox{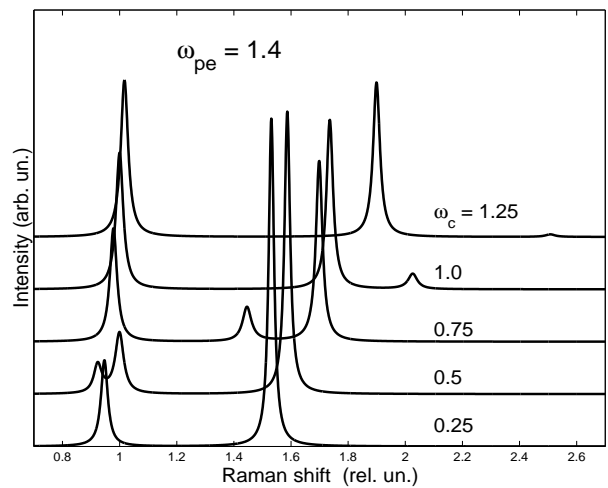}
\end{center}
\caption{
Calculated Raman spectra for the carrier plasma frequency
and cyclotron frequency indicated in figure. All the
frequencies  are in units $\omega_{\rm TO}$; the phonon width
$\Gamma=3\times 10^{-2}\omega_{\rm TO}$, the carrier relaxation rate
$\tau^{-1}=2\times 10^{-2}\omega_{\rm TO}$,  the plasma frequency
of ions $\omega_{pi}=0.4\omega_{\rm TO}$, and
 the momentum transfer $kv_F= 0.3\omega_{\rm TO}$. The cyclotron frequency is supposed
 to be in the semiclassical regime.}
\end{figure}

With the help of Eqs. (\ref{8}) and (\ref{suc}),
we can plot the theoretical Raman spectra for the large,
Fig. 3 and low, Fig. 4 electron concentrations.
The Faust-Henry coefficient is taken $C = - 0.5$ as
is usually accepted in the absence of the magnetic field.
We see always three peaks: the weak resonance approximately
at $\omega=2\omega_c$
and two resonances corresponding $\omega_{\pm}$.
Far from the crossing, the modes have  mainly a
character  of the phonon or plasmon vibrations.

From the splitting in Figs. 1 and 2, 
we can conclude
that the resonance at
$\omega=2\omega_c$ interacts more intensively with the
mode of the plasmon character ($\omega_{+}$ for the large
 concentration of carriers and $\omega_{-}$ for the low concentration).

\begin{figure}
\begin{center}
\epsfxsize=80mm
\epsfysize=64mm
\epsfbox{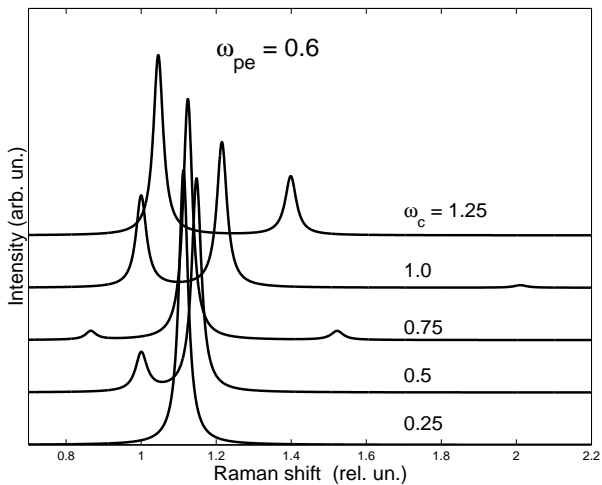}
\end{center}
\caption{
The same as in Fig. 3 but for the carrier plasma frequency indicated in
figure,
$\tau^{-1}=10^{-2}\omega_{TO}$, and for the ultra-quantum regime. }
\end{figure}

The figure 3 (see, also, Fig. 1) clearly  demonstrates the result of
screening in the case of a relatively large concentration of
carriers: the frequency $\omega_{-}$, which equals the frequency
of the longitudinal mode in the absence of carriers (at the
particular choice of parameters,
$\omega_{\rm LO}=1.1\omega_{\rm TO}$),
turns out to be close to the frequency of the transverse mode
$\omega_{\rm TO}$.
The case of the relatively low carrier concentration is
illustrated by Figs. 2 and 4. It is seen in Fig. 4 that, in a weak
field $(\omega_c=0.25\omega_{\rm TO})$, the frequency of the longitudinal mode is
close (as it must be) to $\omega_{\rm LO}=1.1\omega_{\rm TO}$. Besides the
weak resonance at $\omega = 2\omega_c$, a phonon-plasmon crossover
is seen in the curves for $\omega_c=0.75\omega_{\rm TO}~
\mbox {\rm and}~ 1.0\omega_{\rm TO}$.
As the magnetic field increases, the weaker plasmon peak, after
passing through the phonon one, becomes more intense.

\section{Conclusions}

In this work we investigated the   spatial dispersion effect
in the Raman scattering on the phonon-plasmon coupled modes.
In spite of the small value of corresponding parameters
$kv_F/\omega$ (for the semiclassical magnetic fields) and
$ka_H$ (for the ultra-quantum regime), this effect is
observable in the high magnetic fields when the  cyclotron
resonances intersect the frequencies   of the phonon-plasmon
coupled modes. The magnetic field  provides an additional
parameter which gives  a possibility to compare the effect
of  the electron interaction with both the
longitudinal and transverse optical modes
in doped semiconductors.

\begin{acknowledgments}
I am grateful to W. Knap, J. Camassel, and M. Potemski for  discussion of the
work. This work was supported by the Russian Foundation for Basic
Research.
\end{acknowledgments}

\end{document}